\documentclass[times, twoside]{henriques-style}
\usepackage{blindtext}
\usepackage[nolist]{acronym}
\setcitestyle{square}
\usepackage{tikz}
\usetikzlibrary{positioning, calc, matrix, decorations.pathreplacing, shapes.geometric}
\usepackage{adjustbox}
\usepackage{makecell}

\usepackage{xspace}
\makeatletter
\DeclareRobustCommand\onedot{\futurelet\@let@token\@onedot}
\def\@onedot{\ifx\@let@token.\else.\null\fi\xspace}

\leadauthor{Sen} 
\usepackage{color}

\begin{document}

\title{Sk-Unet Model with Fourier Domain for Mitosis Detection
\thanks{$*$ Sen Yang and Feng Luo are co-first authors.}}
\shorttitle{Sk-Unet Model with Fourier Domain for Mitosis Detection}
\author[1,2*]{Sen Yang}
\author[3*]{Feng Luo}
\author[2]{Jun Zhang}
\author[4]{Xiyue Wang}

\affil[1]{College of Biomedical Engineering, Sichuan University, Chengdu, China}
\affil[2]{Tencent AI Lab, Shenzhen, China}
\affil[3]{Shenzhen International Graduate School, Tsinghua University, Shenzhen, China}
\affil[4]{College of Computer Science, Sichuan University, Chengdu 610065, China}

\maketitle

\begin{abstract}
\acl{mc} is the most important morphological feature of breast cancer grading. Many deep learning-based methods have been proposed but suffer from domain shift. In this work, we construct a Fourier-based segmentation model for mitosis detection to address the problem. Swapping the low-frequency spectrum of source and target images is shown effective to alleviate the discrepancy between different scanners. Our Fourier-based segmentation method can achieve F$_1$ with 0.7456, recall with 0.8072 and precision with 0.6943 with on the preliminary test set. Besides, our method reached the 1st place in the MICCAI 2021 MIDOG challenge.

\end {abstract}


\begin{corrauthor}
wangxiyue@stu.scu.edu.cn
\end{corrauthor}

\section*{Introduction}
Nowadays, breast cancer is an increasingly common disease in both developed and developing countries\cite{saha2017advanced}. According to the Nottingham grading system\cite{elston1991pathological}, it can be diagnosed and predicted by three features, which are nuclear polymorphism, mitotic count, and tubule formation on histopathological sections stained with hematoxylin and eosin(H\&E). Among them, mitotic count is the most important morphological feature of grading. So pathologists usually search for mitosis in a complete slide with a high-power fields of view (HPF) manually to count. However, a large number of HPF in a single complete slide and the appearance difference of mitotic cells make the task time-consuming and tedious. In addition, it is objective to judge mitotic cell and are prone to reach a consensus on mitotic count among pathologists.

Recent advances in deep learning and digital scans have paved the way and many automatic mitosis detection methods have been proposed\cite{li2018deepmitosis}\cite{li2019weakly}\cite{sebai2020maskmitosis}. Although achieving great success, a drop in performance is often observed when the trained model is tested on data from another domain(i.e. different slide scanners and sample preparation from clinical centers). This problem makes it hard for mitosis detection algorithms to be widely used in real diagnosis process.

To solve the problem, we construct a Fourier-based segmentation method for mitosis detection and submit it to the MIDOG challenge. We convert mitosis detection task to mitotic cell segmentation, which makes our model more robust and stable. Inspired by \cite{yang2020fda}, We swap the low-frequency spectrum of source and target images to alleviate the discrepancy between different scanners. Experimental results show that our Fourier-based segmentation method can address the domain shift in mitosis detection. It achieves F$_1$ with 0.7456, recall with 0.8072 and precision with 0.6943 with on the preliminary test set of MIDOG challenge.

\section*{Methodology}
Regarding mitosis detection as segmentation, the proposed algorithm can be divided into image pre-processing, fourier domain SK-Unet and image post-processing. Image pre-processing and image post-processing are processes of converting bounding boxes and masks. As for the network, SK-Unet\cite{wang2021sk} equipped with fourier domain adaption is modified for the task.

\subsection*{Image Pre-processing}
Due to the fact that segmentation model is more robust, we convert mitotic detection to segmentation, thus masks of mitotic cells are required. First, all cells in an image are segmented with pre-trained Hovernet\cite{graham2019hover} which is publicly available \footnote{https://github.com/simongraham/hovernet\_inference}. Then we get cells which need to reserve according bounding boxes of the image. In specific, a cell is reserved when the Intersection of Union (IOU) of the cell and any bounding box is over 0.8. 

\subsection*{Fourier Domain Sk-Unet}
\begin{figure}[!ht]
\centerline{\includegraphics[scale=0.3, angle=-90, trim=160 40 160 40, clip]{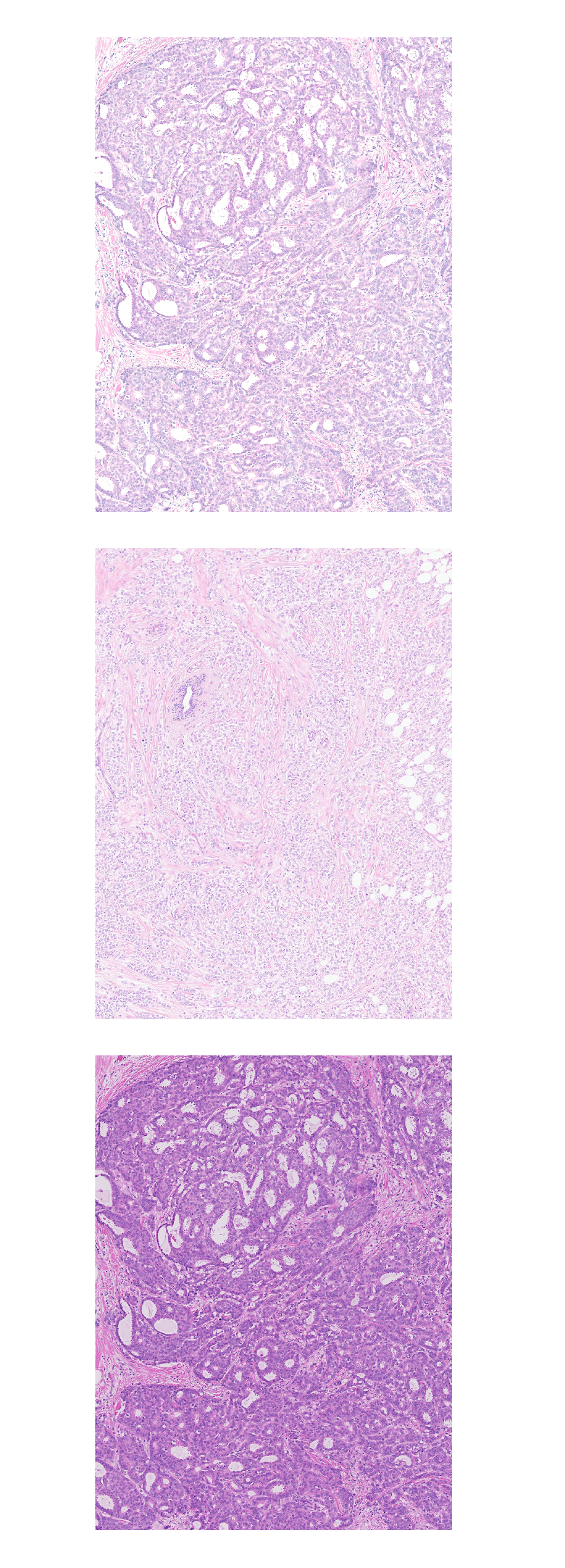}}
\caption{A FDA sample. Images are the source image, reference image and generated image from left to right.}
\label{fda}
\end{figure}

In order to solve the problem of domain adaptation, a simple method for unsupervised domain adaptation is adopted, which is swapping the low-frequency spectrum of one with the other\cite{yang2020fda}. To be specific, there are three steps. First, given an image $I_s$, its amplitude and phase components can be calculated using FFT algorithm\cite{frigo1998fftw}. Second, the center region of $I_s$'s amplitude component is replaced by that of another image $I_t$. This means that low-frequency information of the two images is swapped. Third, the modified amplitude component and its unaltered phase component are used to reconstitute an image with similar style of $I_t$ using inverse FFT (iFFT). The motivation of swapping process is that high-level semantics represented by high-frequency spectrum is the real cue for mitosis while low-level semantics is closer to background information. So combining one high-frequency spectrum with several low-frequency components can generated images with different styles and the same label, which enlarges the amount of training data and enhances the generalization ability of our model. Some generated samples are shown in Fig.~\ref{fda}. For the mitosis segmentation, SK-Unet is adopted by us. The method proposed a combination of feature maps from different scales in the encoder-decoder network to improve the segmentation results.


\subsection*{Post-processing}
The image post-processing process aims to refine the result of cell segmentation and convert it to bounding boxes. Initially, the hole filling technique is applied to attain accurate segmentation masks. Then, connected component analysis for all the obtained masks is performed and each connected component is regarded as a cell. Last, centers of all minimum bounding rectangles for connected components are calculated as our final result.

\section*{Experiment}
\subsection*{Dataset}
Our algorithm is evaluated on the MICCAI 2021 MIDOG challenge\cite{midog}. The MIDOG training subset consists of 200 Whole Slide Images (WSIs) from human breast cancer tissue with four slide scanning systems (Hamamatsu XR NanoZoomer 2.0, the Hamamatsu S360, the Aperio ScanScope CS2 and the Leica GT450). Each scanner has 50 images annotated, except for the Leica GT450. To validate the model, we randomly select 50 images from one of the scanners and model selection is based on models' performance on this validation set. The rest of images were used to train the model. In addition, there is a preliminary test set from MIDOG challenge to evaluate the prior model. It contains 20 images, which are from 2 scanners in the training set and 2 unknown scanners.

\subsection*{Experiment Setup}
A sliding window scheme with overlap is used to crop each WSI into small patches of size 512x512 pixels. Standard real-time data augmentation methods such as horizontal flipping, vertical flipping, random rescaling, random cropping, and random rotation are performed to make the model invariant to geometric perturbations. Moreover, RandomHSV is also adopted to randomly change the hue, saturation, and value of images in the huesaturation-value (HSV) color space, making the model robust to color perturbations.
The Adam optimizer \cite{kingma2014adam} is used as the optimization method for model training. The initial learning rate is set to 0.0003, and reduced by a factor of 10 at the 30th and the 50th epoch, with a total of 80 training epochs. The min-batch size is set as 24. The network is trained by minimizing a total loss function composed of a Focal Loss and a Dice loss. All models are implemented using the PyTorch framework \cite{paszke2017automatic} and all experiments are performed on a workstation equipped with an Intel(R) Xeon(R) E5-2680 v4 2.40GHz CPU and four 32 GB memory NVIDIA Tesla V100 GPU cards.

\subsection*{Experiment Results}
Performance of three models are reported in Table ~\ref{results}. The first two rows are results of LinkNet\cite{chaurasia2017linknet} and SK-Unet\cite{wang2021sk} respectively. It shows the SK module in SK-Unet indeed gets more more informative feature maps in both spatial and channel-wise space than LinkNet. Comparing the second and third row, our model(SK-Unet+FDA) achieves stronger performance on both validation set and preliminary test set. It outperforms SK-Unet 0.0141 and 0.0111 of F1-score, which indicates that FDA (Fourier Domain Adaption) enhances the generalization ability of our model.
\setlength{\abovecaptionskip}{0.cm}
\setlength{\belowcaptionskip}{-0.4cm}
\begin{table}[htbp]
\caption{F1-score from models}
\begin{center}
\begin{tabular}{c|c|c}
\hline
\textbf{Model} & \textbf{\makecell{F1\\(validation set)}} & \textbf{\makecell{F1\\(preliminary test set)}}\\
\hline
\textbf{LinkNet} & 0.6954 & / \\
\textbf{SK-Unet} & 0.7331 & 0.7354\\
\textbf{ours} & 0.7472 & 0.7465\\
\hline
\end{tabular}
\label{results}
\end{center}
\end{table}

\section*{Bibliography}
\bibliography{bibliography}

\begin{acronym}
\acro{mc}[MC] {Mitotic Count}
\acro{midog}[MIDOG]{MItosis DOmain Generalization}
\acro{miccai}[MICCAI]{Medical Image Computing and Computer Assisted Intervention}
\acro{wsi}[WSI]{Whole Slide Image}
\acro{he}[H\&E]{Hematoxylin \& Eosin}
\acro{grl}[GRL]{Gradient Reverse Layer}
\acro{aucpr}[AUCPR]{area under the precision-recall curve}
\end{acronym}

\end{document}